\documentclass[%
 reprint,
 amsmath,amssymb,
 aps,
]{revtex4-2}

\usepackage{graphicx}
\usepackage{dcolumn}
\usepackage{bm}

\begin{document}

\preprint{arXiv:2410.17429}

\title{Exploring Particle Production and Thermal-Like Behavior through Quantum Entanglement}

\author{Alek Hutson}
 \email{ahutson@lamar.edu}
 \altaffiliation[Also at ]{Lamar University}
\author{Rene Bellwied}%
 \email{rbellwie@central.uh.edu}
\affiliation{%
 University of Houston, Physics Department, Houston, TX 77204, USA
}%


\date{\today}

\begin{abstract}
Recent studies have shown a potential correlation between the entanglement of initial state partons in elementary particle collisions, as conceptualized by contemporary quantum and particle theory, and the final state multiplicity distribution of hadrons produced in experiments like those at the Large Hadron Collider (LHC). It has been proposed that this relation between states can be demonstrated in a measurement of entropy. By showing equality between entanglement entropy in the initial state and thermodynamic entropy in the final state, we hope to demonstrate that not only is entanglement the driving mechanism behind matter generation, but also the thermal-like behavior seen in high energy particle collisions.
\end{abstract}

\maketitle


\section{Introduction}
\label{introduction}

The exploration of nuclear structure, phases of strongly interacting matter, and the genesis and distribution of matter across different phase spaces has been significantly advanced by the use of particle colliders, most notably the Large Hadron Collider (LHC). In predicting the behaviour of high energy particle collisions, relativistic hydrodynamics and thermal models have emerged as pivotal tools, that trace the evolution of collisions from an initially thermalized system to the final hadronization step, even for elementary particle collisions [\cite{Becattini:2010sk},\cite{Mazeliauskas:2019ifr}]. While thermodynamic models have successfully predicted particle yields with statistically significant accuracy, they rely on the assumption that the system reaches thermal equilibrium at very early times (less than 1 fm/c). In a thermal framework, equilibrium is achieved through particle interactions, which require time to occur, making it highly improbable that this equilibrium is established at such early times through particle interactions alone. Furthermore, this phenomenological approach fails to capture all the quantum dynamics of the system especially in the initial stages, preventing physicists from gaining a complete description of the systems evolution. All modern Monte Carlo based event generators, such as PYTHIA, describe an elementary proton-proton collision using parton-parton collisions, thus ignoring the impact the non-interacting partons, described by the proton wave function, might have on the reduced density matrix that captures the number of available final states.  
These deficiencies in the current models have led physicists to take new approaches relying on first principles of quantum mechanics to give a more accurate and complete description.

A promising avenue lies in the application of quantum information theory, particularly the concept of entanglement, to elucidate the initial state of colliding systems [\cite{Kharzeev:2017qzs,Bellwied:2018gck,Tu:2019ouv}]. These approaches are not only grounded in fundamental quantum principles, but also supported by empirical evidence found in experiments.
One such experiment conducted by Harvard researchers in 2016 demonstrated that a Bose-Einstein condensate of initially entangled Rb atoms reached quantum thermalization of the neighboring atoms through entanglement after a quench on the system was introduced [\cite{Kaufman:2016mif}]. The system evolved without any particle interactions but reached this thermal-like state instantly, given the inherent instantaneous nature of entanglement. Follow-up experiments, in particular the one by [\cite{Kong:2020}], demonstrated that quantum thermalization through entanglement can be achieved at much higher system temperature and over a larger volume than initially shown.
Similar behaviour in a high energy collisions could therefore explain the seemingly instant emergence of thermal like behaviour in a finite system [\cite{Mazeliauskas:2019ifr}]. This idea was recently confirmed by analyzing the proportionality between the effective temperature of the source and the hard scale of the collision, which can be attributed to the clustering of color sources in the context of initial state entanglement, not only in proton-proton, but also in Pb-Pb collisions [\cite{Feal:2018ptp}].

The intersection of information theory and physics, particularly through Shannon's formulation of information entropy [\cite{Shannon:1948dpw}], has furnished a robust framework for understanding information dynamics, paralleling thermodynamic entropy in its depiction of system disorder. This analogy extends to quantum mechanics, where entanglement epitomizes the quintessential quantum phenomenon, manifesting from superposition and uncertainty principles. It would then stand to reason that by studying the quantum coherent properties of a system we can derive the macroscopic behavior described by statistical thermodynamics. 

In quantum mechanics, the state of a system is often represented by density matrices, particularly for composite systems. The elements of a density matrix encode the probability amplitudes of the system's states, thereby capturing all information about the system. The degree of entanglement, or the information shared between a subsystem and the remainder of the system, can be quantified using the von Neumann entropy, defined for a subsystem \( A \) as:
\begin{equation}
S(A) = -\mathrm{Tr}(\rho_A \ln(\rho_A)),
\end{equation}
where \( \rho_A \) is the reduced density matrix of subsystem \( A \). For a pure state, described by a single wavefunction, the von Neumann entropy of the complementary subsystem (the remainder of the system outside \( A \)) is equal in magnitude and opposite in sign, resulting in a total entropy of zero for the composite system. This entropy also known as entanglement entropy quantifies the level of disorder in the system and provides a quantum analog to thermodynamic entropy.

The initial state of a proton-proton collision, viewed through the parton model, presents a complex interplay of valence quarks and a gluonic sea, further complicated by gluon-mediated quark-antiquark pair creation. This pre-collision state, potentially describable by a single wavefunction, suggests a high degree of initial entanglement among the partons.

The entanglement of partons in protons and the subsequent evolution during collisions offer a novel lens to examine high-energy physics phenomena. By quantifying entanglement in the initial state and comparing it to the entropy in the final state, we aim to unravel the role of quantum information processes in particle collisions, potentially bridging the gap between quantum mechanics and thermodynamic behaviors observed in high-energy physics experiments.

\section{Defining the Initial State}
In pp collisions, entropy in the initially pure system originates from a break in entanglement between a probed and a non-interacting region, see Fig.1. 

\begin{figure}[hbtp]
\centering
\includegraphics[width=.45  \textwidth]{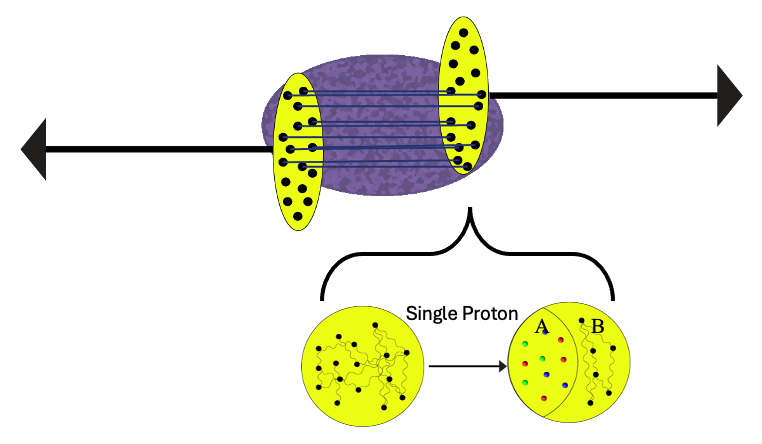}
\caption{Entropy is generated in the transverse direction due to decoherence between the entangled partons probed in the interaction region A and the remainder of the partons in region B. String generation in the longitudinal direction provides a third dimension to the interacting system thereby making the entropy an extensive quantity.}
\label{fig-1}       
\end{figure}

By tracing over the degrees of freedom in the reduced density matrix of the system, one can calculate the entanglement entropy between different regions. This calculation can be challenging due to the infinite and complex nature of the density matrix, driven by the numerous partonic degrees of freedom. However, in the limit of small momentum fractions \( x \), where the proton is described as a highly dense system of indistinguishable gluons, the relevant degrees of freedom can be effectively reduced to the parton number. In this scenario, the density matrix simplifies to:
\begin{equation}
\rho_A \approx \frac{1}{N} I_{N \times N}
\end{equation}
where \( N \) represents the number of partons, and $I_{N \times N}$ represents the identity matrix with NxN dimensions. Consequently, by restricting the kinematics of our collision system to low \( x \), we can simplify the calculation of the entanglement entropy, which reduces to:
\begin{equation}
S(A) \approx \ln(N)
\end{equation}
Here, \( N \) is determined by integrating over the known parton probability distribution, with integration limits based on the spatial region measured in the final state. PDF's are provided in terms of the parton momentum fraction or Bjorken-x and momentum transfer  Q$^{2}$. These kinematics are set by the energy of the collision and the magnitude of the overlap region respectively. 
The relevant momentum transfer scale Q$^2$ is known as the saturation scale and can be related to x through the Balitsky–Kovchegov equation. In p-p collisions we can not directly measure Q$^2$ therefore we consider the average value of Q$^2$ based on the average value of x. The data analysis presented here is restricted to the central rapidity region covered by the Time Projection Chamber (TPC) in the ALICE detector at the LHC, which limits our x-range to values around 10$^{-4}$, corresponding to a Q$^{2}$-range around unity [\cite{Albacete:2012xq}, \cite{Casuga:2023dcf}].

It is important to note here that in this low Q$^{2}$ regime higher order expansions of the QCD coupling constant give unreliable results for gluon distributions. Therefore we only consider leading order (LO) parton distribution functions (PDFs). Even here, the uncertainties, in particular in the gluon distributions are sizeable. As examples we show the LO results from the NNPDF [\cite{Ball:2011uy}] and MHST [\cite{Bailey:2020ooq}] collaborations in Figs.2 and 3, respectively. The large error bars in the nuclear PDFs might be further constrained, by applying entanglement arguments to the particle multiplicity measurements.

\subsection{Entanglement of collision stages}
Unitarity of the S matrix in quantum mechanics tells us that an entangled quantum system will have no net increase in entropy as long as there is no environment to which information can be lost. In a p-p interaction we can treat one proton as the system in question and the other proton as the environment to which information is lost. This initial interaction sets the entropy of the interacting system.

Prior to the collision we have a Lorentz contracted proton moving near the speed of light. Based on the parton model it exists as a set of semi-free partons (gluons and quarks) in a flat pancake-like shape. In reality these partons exist in a complete coherent state meaning that the constituents are completely entangled. The purely coherent state of the proton contains zero von-Neumann entropy. This can be understood using the protons density matrix. The density matrix of a particle contains all the information about that particle including its spin states, position, momentum etc. Because this matrix contains all information in every Hilbert space the particles density matrix will end up being infinite in its dimensionality.

To simplify and illustrate our point let us consider the density matrices that describe the proton in the spin Hilbert space. Such a density matrix would be expressed as follows:
\begin{equation}
    \rho = \frac{1}{2} (\mathbb{I} + \vec{r} \cdot \vec{\sigma})
\end{equation}
Where $\sigma$ represents the Pauli matrices for three spatial dimensions: 
\begin{equation}
\sigma_x = \begin{pmatrix} 0 & 1 \\ 1 & 0 \end{pmatrix}, \quad \sigma_y = \begin{pmatrix} 0 & -i \\ i & 0 \end{pmatrix}, \quad \sigma_z = \begin{pmatrix} 1 & 0 \\ 0 & -1 \end{pmatrix}
\end{equation}

\begin{figure}
\centering
\includegraphics[width=.45\textwidth]{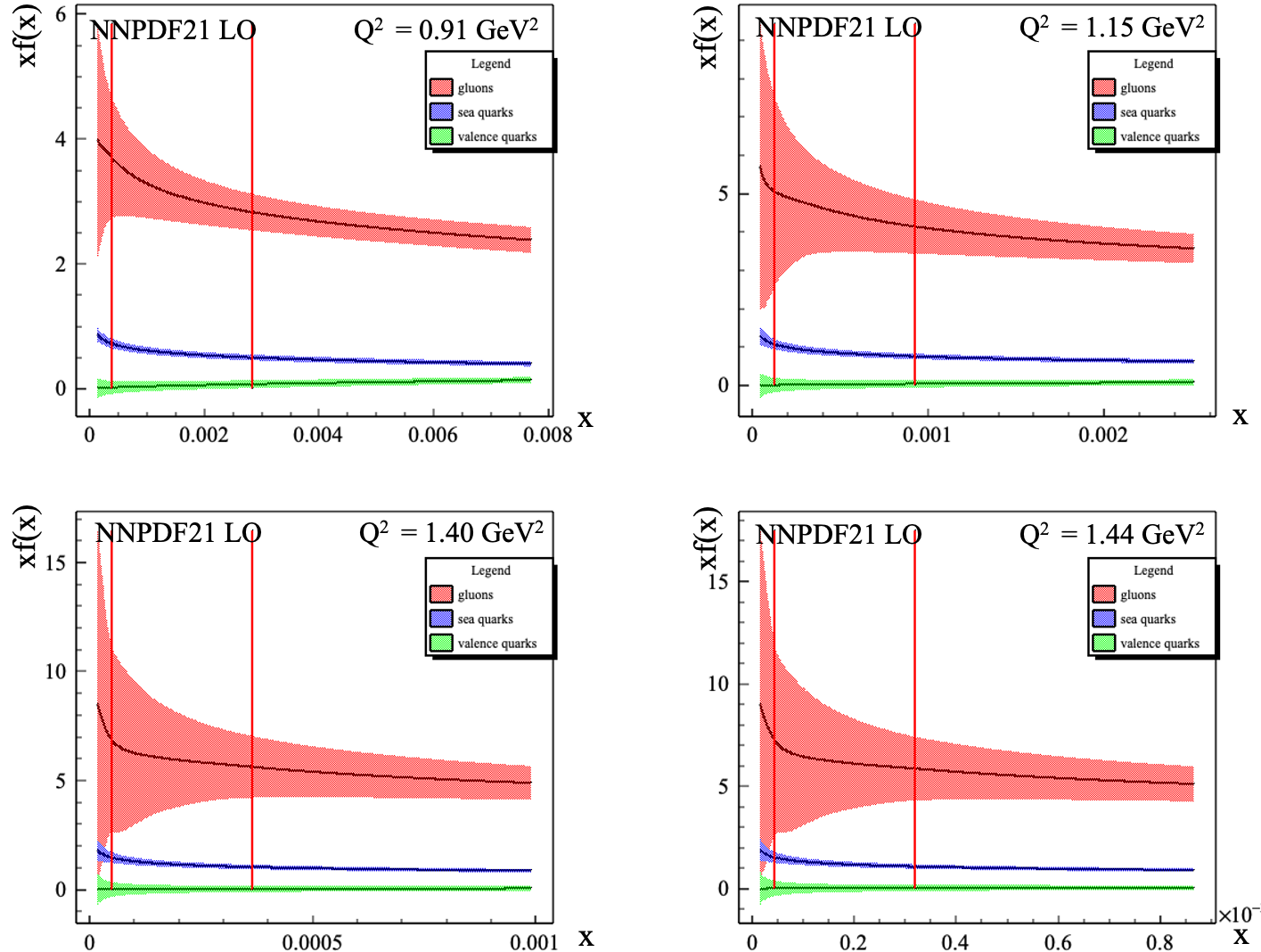}
\caption{ PDF sets defining the initial state of the proton extrapolated from NNPDF collaboration data sets at a reference scale $\alpha_s(M_z)$ = 0.119 [\cite{Ball:2011uy}]. PDF's were extrapolated using using LHAPDF [\cite{Buckley_2015}].  }
\label{fig-2}       
\end{figure}

\begin{figure}
\centering
\includegraphics[width=.45\textwidth]{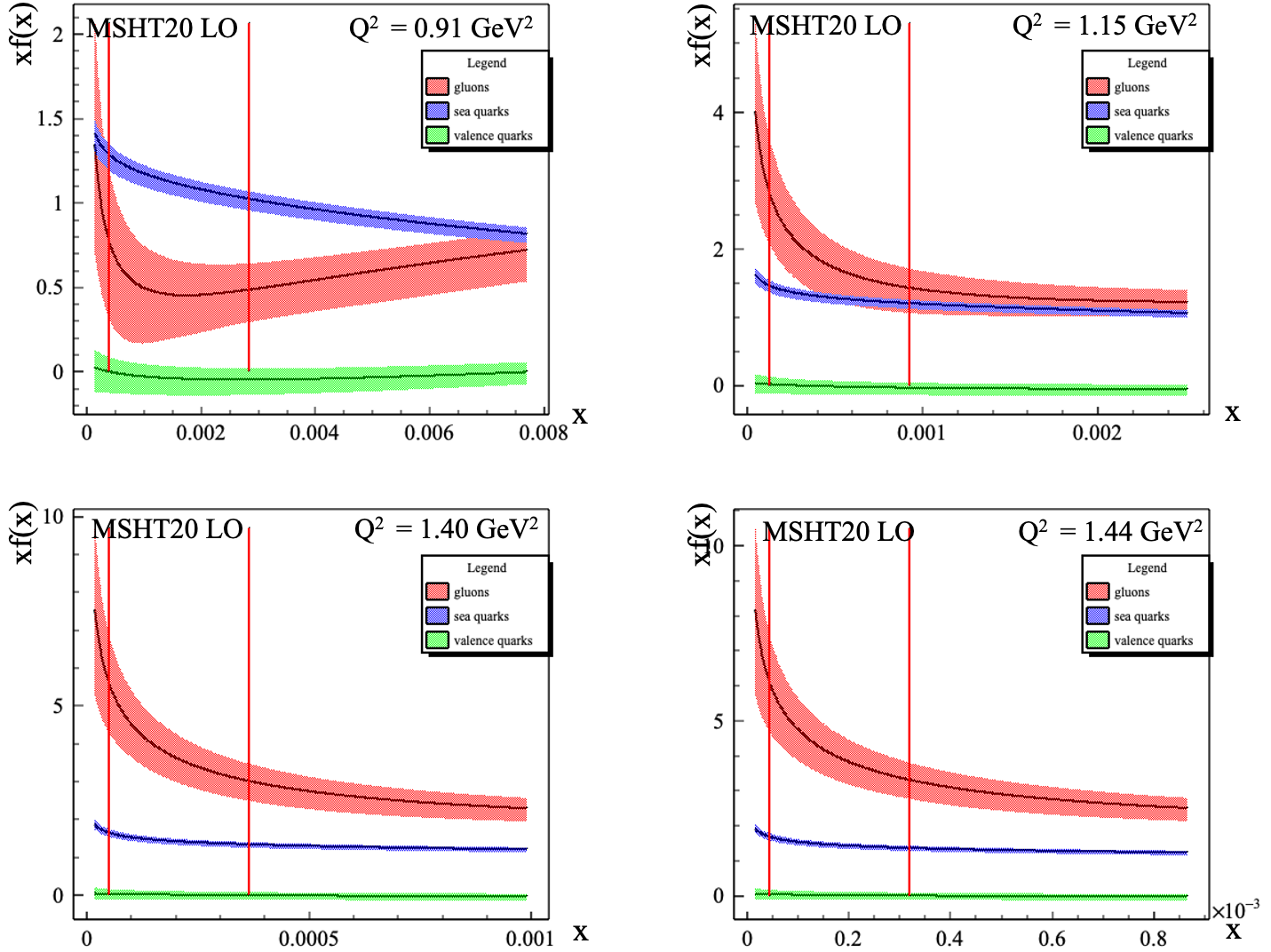}
\caption{ PDF set defining the initial state of the proton extrapolated from MSHT collaboration data sets at a reference scale $\alpha_s(M_z)$ = 0.130 [\cite{Bailey:2020ooq}].  PDF's were extrapolated using using LHAPDF [\cite{Buckley_2015}].}
\label{fig-3}       
\end{figure}

Now if we computed $\rho$ for any given direction and diagonalized the matrix we would find the von-Neumann entropy to be zero. If the system of partons has zero entropy then they must be fully entangled with one another. Furthermore, because the system exists in a pure state quantum mechanics dictates that the global state of the system must remain pure. This purity manifests itself through conservation of quantum numbers.

At the onset of the collision there is a break of entanglement and a loss of information between the overlap region and the remainder of the proton. This information is encoded in the reduced density matrix that defines all quantum states of constituent partons within the overlap region. The distribution of partons in this region is dependent on the kinematics of the collision. The aforementioned parton distribution functions describe the wave function as gluon dominated for regions at low-x. This gluon dominance simplifies the reduced density matrix such that we can estimate it as a collection of indistinguishable gluons which makes the spin, spatial, and other degrees of freedom negligible and requires only the number of partons involved in the initial collision.

During the collision, interacting partons are halted, while non-interacting partons, the environment to which entanglement is lost, continue to carry momentum forward. This process results in an isolated system of interacting partons, which retains some initial non-zero entropy. 

After the initial quench of the wave function the combined system of interacting partons interacts through flux tube or string generation along the beam axis. This string generation adds a third dimension to the initially two dimensional picture, thereby generating an extensive volume [\cite{Berges:2017hne}]. In theory these flux tubes can then merge creating a coherent initial state described as a color glass condensate0 [\cite{Gelis:2010nm}]. Because the subsequent system exists in this coherent state the resultant longitudinal entanglement would yield negligible entropy generation. Therefore the total entropy generated is equal to the entropy generated at the ends of the strings and arises from decoherence between the interaction region and the non interacting region of the two-dimensional Lorentz contracted proton. Furthermore, after the collision the interacting system is now in isolation from any external environment with which to exchange information, so the von Neumann entropy remains constant. 

\subsection{Modeling Entanglement Entropy}

We seek to establish an equivalence between the initial-state entanglement entropy, calculated using Parton Distribution Functions (PDFs), and the final-state entropy derived from multiplicity distributions of primary charged particles, as measured by the ALICE detector. Hints of this equivalency have been seen in e-p interactions [\cite{Hentschinski:2022rsa}], but in proton-proton collisions the overlapping of two proton wave functions adds complexity to the problem. Starting from equation 3 we can begin to calculate an entropy in the low-x limit based on the number of interacting gluons calculated by summing over PDF's:

\begin{equation}
N_{\text{gluons}} = \sum_{x=\text{xmin}}^{\text{xmax}} \frac{xf(x)}{x} \Delta x
\end{equation}

The need to extend this formalism to quark contributions, in particular sea quark contributions at low Q$^{2}$ was shown by Hentschinski et al. [\cite{Hentschinski:2022rsa}], and their proposed formalism 

\begin{equation}
N_{\text{partons}} = \sum_{x=\text{xmin}}^{\text{xmax}} \frac{xf_{gluon}(x)}{x} \Delta x + \sum_{x=\text{xmin}}^{\text{xmax}} \frac{xf_{quark}(x)}{x} \Delta x
\end{equation}

has been used for the final comparison to data. Hentschinski et al. also add a factor to this calculation to account for unmeasured neutral particles in the final state, since only charged particle multiplicity distributions are measured. To account for this loss we assume that one-third of the initial state partons are used to produce neutral particles. Therefore when calculating N for the initial state entropy measurement we multiply by a factor two-thirds to account for only particles which produce positively charged and negatively charged hadrons. Equation (6) then becomes:
\begin{equation}
    S_{EE} = ln(2/3(N_{gluons} + N_{quarks})) = ln(N_{gluons} + N_{quarks}) + ln(2/3)
\end{equation}

Finally, entropy measured in the final state is only taking into account the degrees of freedom related to the number of particles produced. However, the actual entanglement entropy is calculated by summing over the infinite density matrix. Constructing such a density matrix would be impossible given the finite resolution and constraints of the detector. In information theory this entropy is known as the entropy of ignorance, because it is only sensitive to the number of produced hadrons and does not consider all the degrees of freedom within the density matrix. Using an initial CGC model theorists have been able to calculate the ratio between this entropy of ignorance (measured entropy) and the entanglement entropy, as shown in Fig.4 [\cite{Duan:2020jkz}]. In the calculation it is observed that at low Q$^{2}$ this ratio is very large and has an upper limit of about 1.4. At high Q$^{2}$ like those seen in DIS measurements this ratio saturates to one, making the discrepancy negligible. For LHC energies we see this ratio reach around 1.24 for $\sqrt{s}$ = 0.9 TeV. We apply this ratio correction to the initial-state entropy in order to correct for the missing information in the final-state distribution.

\begin{figure} [hbtp]
\centering
\includegraphics[width=.40\textwidth]{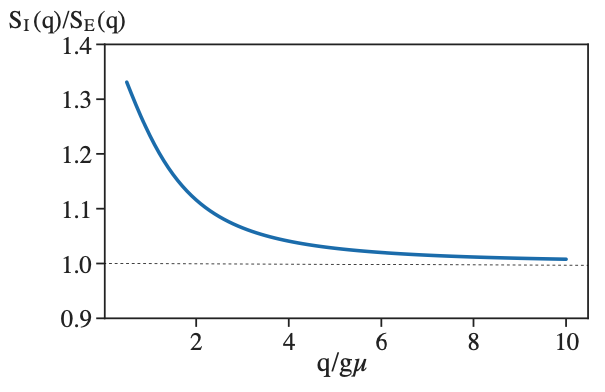}
\caption{ Ratio between the entropy of ignorance and entanglement entropy in the initial state [\cite{Duan:2020jkz}].  }
\label{fig-4}       
\end{figure}

\section{Defining the Final State}
Unitarity requirements of the QCD S-matrix require that the entanglement entropy of the global system remains zero. As much as any quantum number is conserved, so is the quantum information and subsequent entropy. 
\begin{equation}
S_A = -\rho_{A} \ln \rho_{A} = S_B = -\rho_{B} \ln \rho_{B}
\end{equation}
However, due to the infinite nature and complexity of the complete density matrix of states of produced hadrons we are unable to measure a true entanglement entropy, but instead we are able to measure a Shannon entropy which is proportional to the thermodynamic entropy. The thermodynamic entropy is defined in a similar way as the entanglement entropy in as much as it is a  sum over probabilities to quantify the level of disorder in the system. 
\begin{equation}
S_{\text{hadron}} = -\sum P(N) \ln P(N)
\end{equation}

Since the fractional momentum and momentum transfer scale are not directly measurable in proton-proton collisions, an alternative way of comparing entropies has to be employed. In the final-state we can calculate a thermodynamic entropy using the probability distribution of produced particles. By setting pseudo-rapidity boundaries and counting particles in a range we can make a meaningful comparison to the initial state, integrated over the same kinematic range. This range can be any width in $\eta$ as long as we can define a mapping between the pseudo-rapidity in the final state and the fractional momentum \textit{x} in the initial state. Within the CGC model a common approximation is given by  

\begin{equation}
ln (1/x) = y_{\text{proton}} - y_{\text{hadron}}
\end{equation}

This relation might break down at large x [\cite{Jalilian-Marian:2020bwd}], but we are mostly interested in the low x, high gluon density region. For this analysis it is convenient to measure charged particles at mid-rapidity where we have strong particle tracking abilities and high statistics.

ALICE has published charged particle multiplicity data for four collision energies ranging from $\sqrt{s}$ = 0.9 - 8 TeV [\cite{ALICE:2015olq}]. Entropy in the final state is defined by the level of disorder in the multiplicity distribution P(N), which defines the probability of finding N produced hadrons after the collision. The measured multiplicity distributions from p-p collisions at central pseudo-rapidity are shown in Fig.5. The data are well described by a Negative Binomial Distribution (NBD). Since the particle production is generated by the collision system, a transformation is required in order to make a meaningful comparison to PDFs, which represent the parton distribution in a single proton. An assumption is made that half of the produced hadrons are coming from one proton and half is coming from the other [\cite{Tu:2019ouv}]. A fit to half the multiplicity is then used to calculate the Shannon entropy from a single proton. 

\begin{figure}[hbtp]
\centering
\includegraphics[width=.35\textwidth]{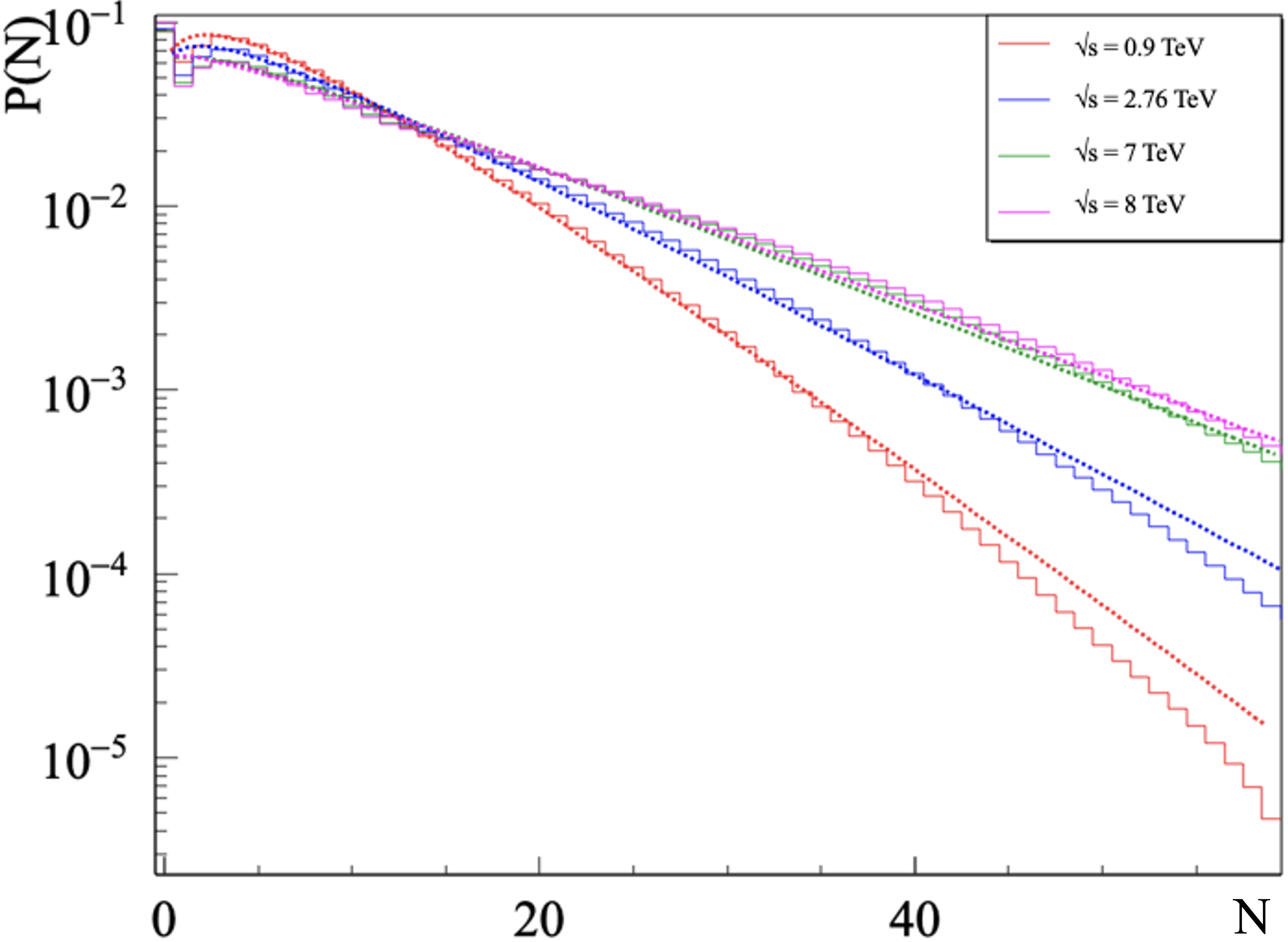}
\caption{Produced hadron multiplicity distributions from ALICE data. [\cite{ALICE:2015olq}]. The dashed lines show a fit using a negative binomial distribution (NBD).}
\label{fig-5}
\end{figure}

Higher moments of the NBD distribution are used to compare with a 1+1 toy model of non-linear QCD evolution of the BK equation as suggested by Kharzeev and Levin [\cite{Kharzeev:2017qzs}]. They demonstrated that one can construct a generating function that captures the non-linear interactions leading to dipole formation, in an entangled system, which sets an upper limit on the moments of the final state NBD distribution. 

\begin{equation}
C_{2} = 2 - \frac{1}{\bar{n}}
\end{equation}

\begin{equation}
C_{3} = \frac{6 (\bar{n} - 1) \bar{n} + 1}{\bar{n}^{2}}
\end{equation}

\begin{equation}
C_{4} = \frac{(12\bar{n} (\bar{n} - 1) + 1)(2\bar{n} - 1)}{\bar{n}^{3}}
\end{equation}

\begin{equation}
C_{5} = \frac{(\bar{n} - 1)(120\bar{n}^{2}(\bar{n} - 1) + 30\bar{n}) + 1}{\bar{n}^{4}}
\end{equation}

Where $\bar{n}$ represents he average multiplicity from data. We observe that the final state multiplicity distribution approach the limit expected from a fully entangled state, see Fig.6.

\begin{figure}[hbtp]
\centering
\includegraphics[width=.45\textwidth]{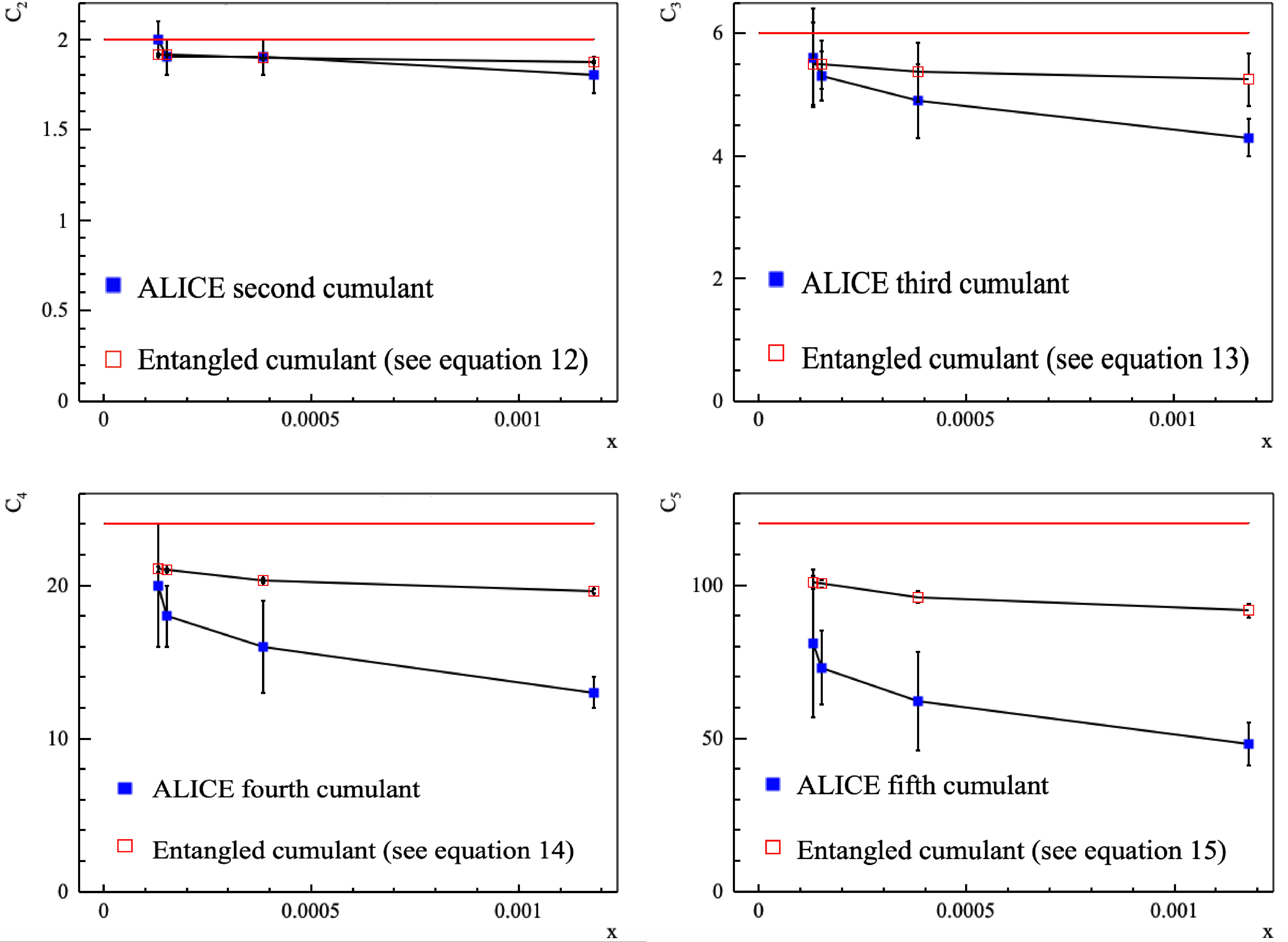}
\caption{ Filled squares show final state cumulants from ALICE distributions [\cite{ALICE:2015olq}]. Open squares show theoretical cumulants of a parton cascade using an entangled model [\cite{Kharzeev:2017qzs}]. Red lines represent the upper bound on entangled cumulants in the limit of $\bar{n}$ approaching infinity.}
\label{fig-6}
\end{figure}

\section{Results and Conclusions}

In the following we show the derived final state thermodynamic entropy (red points) in comparison to various approaches to the initial state entropy, either from partonic string fragmentation or entanglement calculations. Final state entropy is calculated from the published multiplicity data from ALICE for the four points at higher x, and the point at lowest x is calculated by extrapolating these distributions using a power law fit of the mean.

Particle production in proton-proton collisions is often modeled by string fragmentation models, such as PYTHIA, that generally ignore spectator partons in the proton. Even the most recent PYTHIA tunes underestimate the final state entropy significantly when not taking into account some type of quantum effects. Lately these effects have been modeled by two distinct PYTHIA parameters and modes, namely the level of multi-parton interactions (MPI) and the color reconnection mode (CR).  Fig.7 shows the comparison of the final state entropy from ALICE data to various PYTHIA modes. One can see that a fragmentation model that takes into account ad-hoc approaches to quantum mechanically driven interaction and coalescence processes does significantly better than the standard fragmentation approach.

\begin{figure}[hbtp]
\centering
\includegraphics[width=0.45\textwidth]{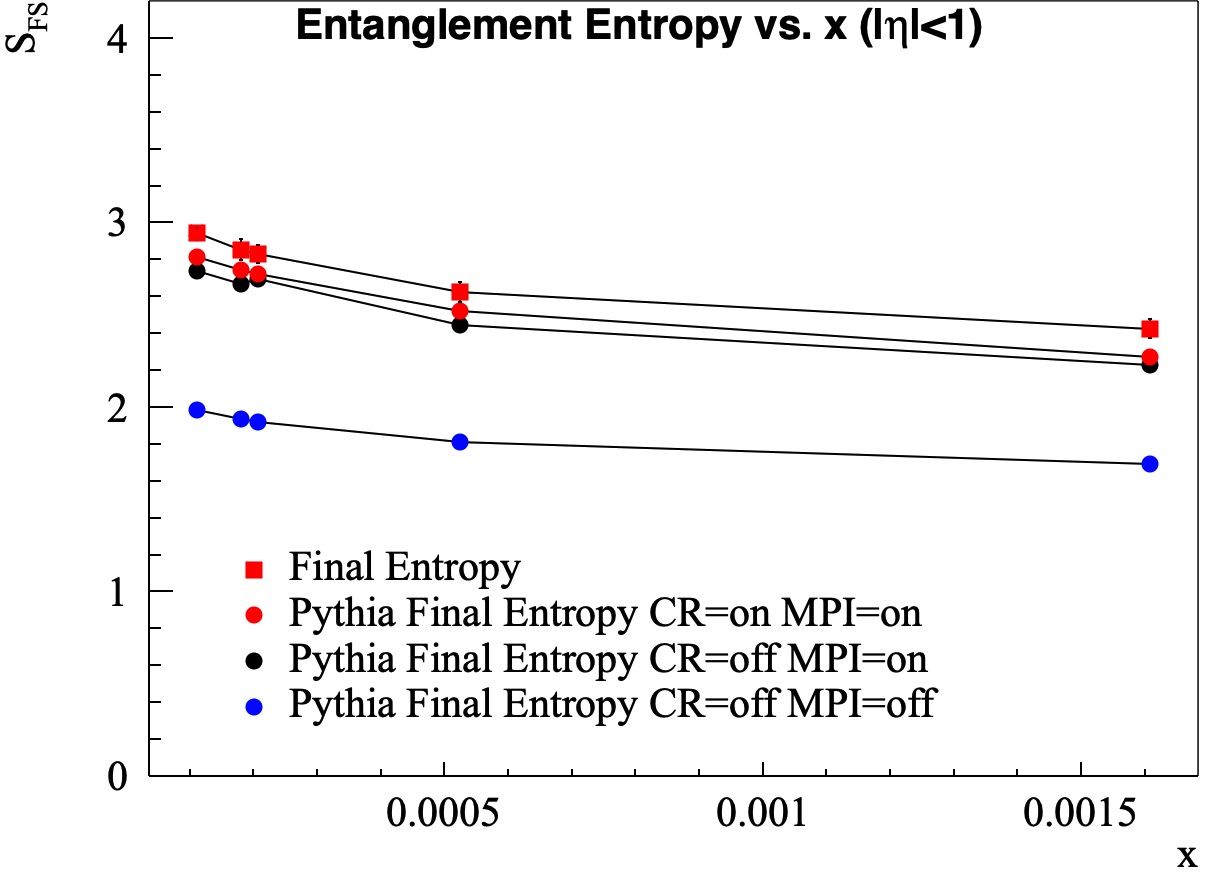}
\caption{Final state entropy calculated from ALICE data compared to three different tunes of PYTHIA. }
\label{fig-7}
\end{figure}

Fig.8 shows the final state entropy compared with several initial-state entropy calculations shown as bands. The red band represents the initial-state only considering gluons. The magenta band represents the initial-state with both gluons and quarks. Finally, the green band represents the initial-state considering gluons and quarks and correcting for the entropy of ignorance (see Fig.4 [\cite{Duan:2020jkz}]).
The final state charged particle multiplicity is used as the absolute measure and only converted to entropy. All correction factors are then applied at the parton level.

\begin{figure}[hbtp]
\centering
\includegraphics[width=0.45\textwidth]{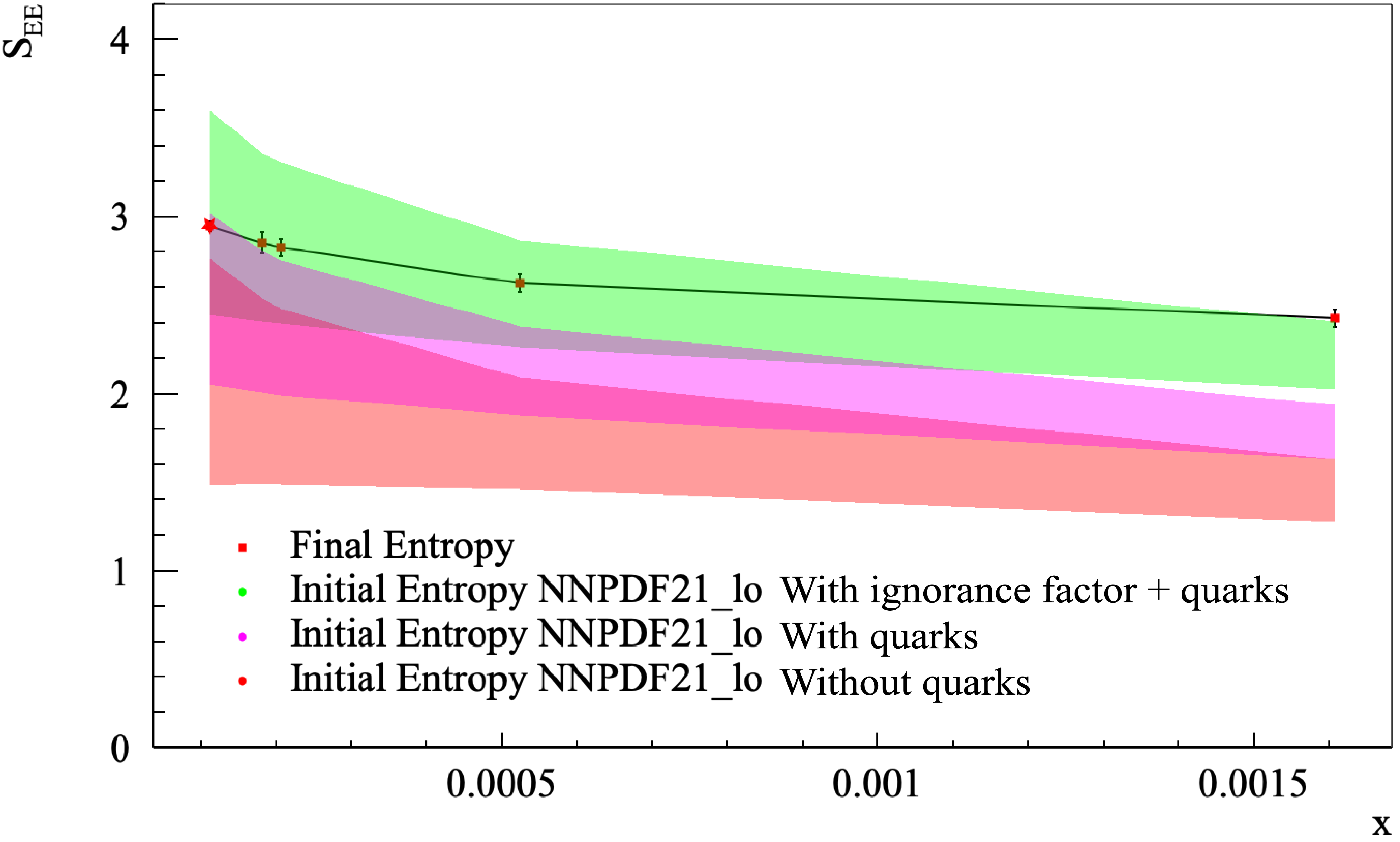}
\caption{ Finale state (red points) compared to initial state entropy under different initial state assumptions.}
\label{fig-8}
\end{figure}

\begin{figure}
\centering
\includegraphics[width=0.45\textwidth]{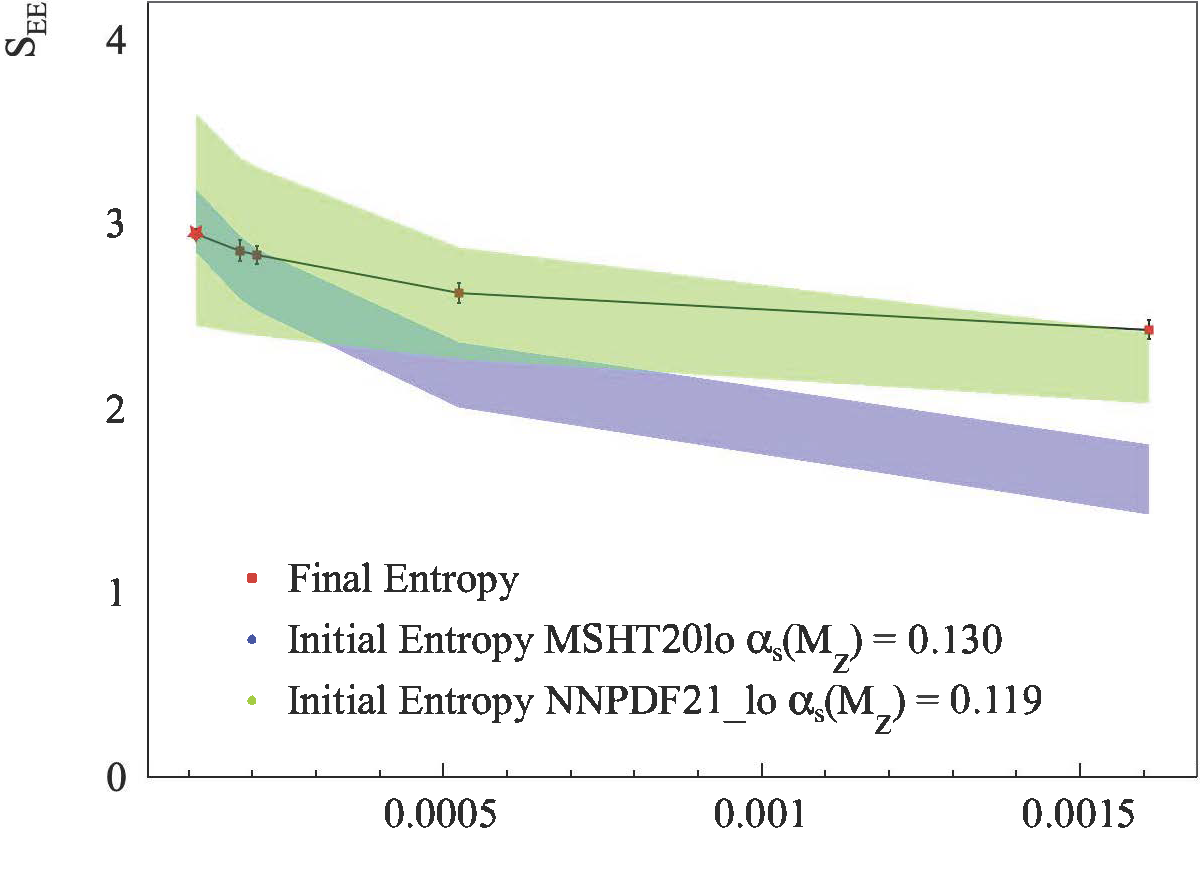}
\caption{ Final state entropy in red, and the initial state entropy shown by colored bands for two models including all partons and corrections.  }
\label{fig-9}
\end{figure}

Fig.9 shows a comparison between two different leading order parton distribution functions at the relevant x, Q$^{2}$ and $\alpha_s$ values.

We have shown that the entropies calculated from hadron multiplicity in the final state and parton multiplicity in the initial state approach one another at low-x, as long as contributions from all partonic states are taken into account. We also observe stronger agreement between the initial state and final state for a coupling constant equal to 0.119 as opposed to 0.130. Our approach thus further verifies the lower coupling constant predicted by recent model calculations. 

We have not seen any effect of gluon saturation in the covered x-range, but future measurements at forward rapidity at the LHC and over the full phase space at the EIC, will be highly relevant, since it is predicted that the entropy generation should follow the gluon saturation curve at very low x.

This paper shows, for the first time, a multi-layered analysis of entanglement effects that lead to a quantitative agreement with LHC based data. On the data side, the ALICE results cover a significantly different phase space than previous analyses, in terms of low momentum and central rapidity coverage. The low Q$^{2}$ region is emphasized which leads to a dataset that is more conducive to gluon entanglement.
On the theoretical side, the impact of the quark distribution functions is not negligible and the entropy of ignorance factor can be analytically confirmed.

Overall our measurements give support to the theory that the quantum entangled state in the initial phase of a relativistic hadron collision could maintain its coherence throughout the early evolution and lead to a seemingly thermal final state, where the observed particle multiplicities are governed by the partition function of the initial parton states.

\begin{acknowledgments}
The work presented here is supported by the U.S. Department of Energy, Office of Science, under Award Number DE-FG02-07ER41521.
\end{acknowledgments}


\bibliography{apssamp}

\end{document}